\documentclass[pdflatex,iicol,sn-mathphys-num]{sn-jnl}

\usepackage{graphicx}%
\usepackage{multirow}%
\usepackage{amsmath,amssymb,amsfonts}%
\usepackage{amsthm}%
\usepackage{mathrsfs}%
\usepackage[title]{appendix}%
\usepackage{xcolor}%
\usepackage{textcomp}%
\usepackage{manyfoot}%
\usepackage{booktabs}%
\usepackage{algorithm}%
\usepackage{algorithmicx}%
\usepackage{algpseudocode}%
\usepackage{listings}%
\usepackage{tikz}%
\newcommand*\circled[1]{\tikz[baseline=(char.base)]{
            \node[shape=circle,draw,inner sep=0.6pt] (char) {#1};}}

\usepackage{chemformula}

\theoremstyle{thmstyleone}%
\theoremstyle{thmstyletwo}%

\theoremstyle{thmstylethree}%

\begin{document}

\title[Article Title]{Agentic programs as an emerging form of scientific software in computational materials science}


\author[1]{\fnm{Yunsung} \sur{Lim}}
\equalcont{These authors contributed equally to this work.}
\author[2]{\fnm{Haekwan} \sur{Jeon}}
\equalcont{These authors contributed equally to this work.}
\author[2]{\fnm{Jaesun} \sur{Kim}}
\author*[1]{\fnm{Jisu} \sur{Kim}}\email{kskjs1203@snu.ac.kr}
\author*[1,2,3]{\fnm{Seungwu} \sur{Han}}\email{hansw@snu.ac.kr}

\affil[1]{\orgdiv{Research Institute of Advanced Materials}, \orgname{Seoul National University}, \orgaddress{\city{Seoul}, \postcode{08826}, \country{Republic of Korea}}}

\affil[2]{\orgdiv{Department of Materials Science and Engineering}, \orgname{Seoul National University}, \orgaddress{\city{Seoul}, \postcode{08826}, \country{Republic of Korea}}}

\affil[3]{\orgdiv{Center for AI and Natural Sciences}, \orgname{Korea Institute for Advanced Study}, \orgaddress{\city{Seoul}, \postcode{02455}, \country{Republic of Korea}}}


\abstract{Computational materials science has traditionally delegated algorithmic tasks to computers while leaving scientific judgments to humans. \textcolor{black}{In this Perspective, we argue that agent harnesses built on large language models (LLMs) enable an emerging form of scientific software, agentic programs, integrating deterministic algorithms with bounded LLM-based judgment, task-specific verification, episodic maturation, and complete delegation. We illustrate these characteristics through DeMARS, which constructs atomistic models from experimentally measured disordered crystal structures.}}



\maketitle

\section{Introduction}

Since its birth, computational materials science has developed with a clear division of labor: humans develop algorithms and codes and carry out analysis, while computers crunch numbers~\cite{vasp, quantum_espresso, lammps, gromacs, raspa}. This division has enabled researchers to delegate well-defined computational tasks to computers and execute them efficiently at scale. Accordingly, in understanding mechanisms and discovering new materials, significant effort has been devoted to developing efficient algorithms and reliable automation packages that fit into computational workflows~\cite{pymatgen, atomate2, amp2, aiida, aflow}. However, many important decisions in research still rely on human reasoning, which remains outside these workflows~\cite{human_reasoning1, human_reasoning2}. If such reasoning could also be incorporated into the computational process, research goals could potentially be reached more efficiently.

\begin{figure}
\centering
  \includegraphics[width=9cm]{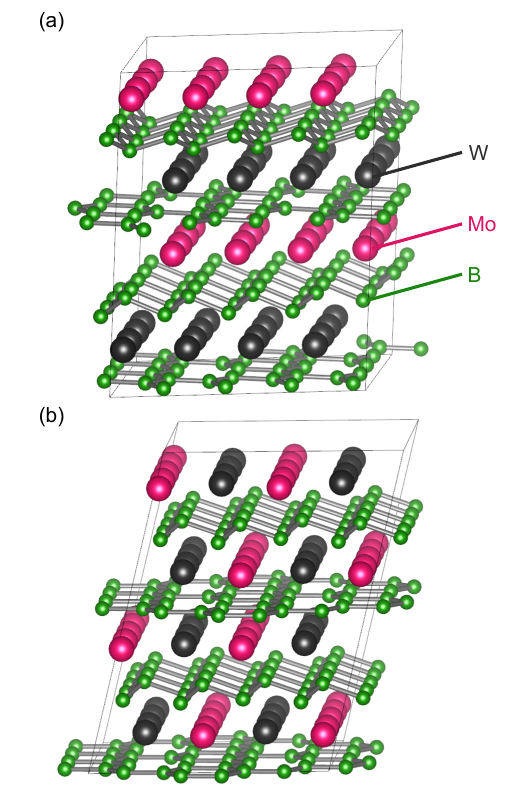}
  \caption{\textbf{Crystal structures predicted for \ch{W4Mo4B15}.} Each structure is obtained by utilizing (a) machine-learning interatomic potential and (b) first-principles calculation to evaluate potential energy used for genetic algorithm.}
  \label{fgr:crystal_structure}
\end{figure}

One example is illustrated in Fig.~\ref{fgr:crystal_structure}, adapted from ref.~\cite{spinner}, which employed a genetic algorithm for crystal structure prediction of the unknown compound \ch{W4Mo4B15}. With a machine-learning interatomic potential (MLIP) enabling a much larger number of search iterations, the algorithm identified a structure in Fig.~\ref{fgr:crystal_structure}a that is more stable than those reported in the previous study~\cite{uspex} (Fig.~\ref{fgr:crystal_structure}b). This demonstrates how improved algorithms and higher computational throughput can expand the search space. Nevertheless, a researcher examining Fig.~\ref{fgr:crystal_structure}b may readily recognize its layered character and reason that different types of intercalation of W and Mo should be explored. Such reasoning can redirect the search toward a promising region of configuration space without relying solely on more extensive sampling.

Although reasoning can accelerate research progress, it has been difficult to incorporate it directly into computational workflows because human reasoning is costly, slow, and difficult to scale~\cite{human_reasoning3}. This situation is changing with the development of large language models (LLMs), which can perform scientific reasoning and translate it into actionable judgment at much higher throughput~\cite{coscientist2023,coscientist2026}. Furthermore, when coupled with an agent harness that can execute computational tools, inspect intermediate results, and act according to its judgment, such reasoning can become an integral part of the computational workflow rather than remaining an external human step~\cite{chemcrow,matsci_agent1}. This opens the possibility of delegating to computers not only numerical calculations but also scientific tasks that require judgment during their execution.

\textcolor{black}{In this Perspective, we argue that agentic programs represent an emerging form of scientific software that extends computational delegation to bounded scientific judgment. We develop this view through four defining characteristics, illustrate their integration through DeMARS, and discuss the challenges and prospects of this form of software.}

\section{Defining characteristics of agentic programs}

We refer to this emerging form of scientific software as an \emph{agentic program}. Unlike an interactive AI assistant~\cite{sciscigpt} or an agent embedded in a human-supervised workflow~\cite{coscientist2026}, an agentic program is designed to take full responsibility for a bounded scientific task. Given a defined input, goal, and set of scientific constraints, it combines conventional algorithms with LLM-based judgment to determine and execute the actions needed to produce a valid output. Importantly, once sufficiently mature, the program is expected to operate without routine human intervention in production, much like conventional scientific software. In this sense, agentic programs extend the long-standing division of labor in computational materials science from the delegation of numerical computation to the delegation of bounded scientific judgment. We suggest that an agentic program has the following defining characteristics.

\textbf{Algorithm and judgment.} An agentic program does not replace conventional algorithms with an LLM. Rather, it combines the two according to their respective strengths. Operations that can be reduced to explicit rules should remain deterministic and algorithmic, while decisions that depend on context, scientific knowledge, or interpretation can be assigned to LLM-based judgment. This hybrid use of algorithms and model-based judgment is shared with emerging agentic workflows in science, where LLMs may gather evidence, generate hypotheses, or determine subsequent actions~\cite{chemcrow, matsci_agent1, coscientist2026}. In an agentic program, however, judgment is not primarily used for open-ended exploration, but for resolving underdetermined decision points within a predefined scientific responsibility. The role of the LLM is therefore to handle those parts of the workflow where the appropriate course of action cannot be completely specified in advance. Importantly, such judgment can have bounded control-flow authority, allowing the program to select among procedures, revise its computational strategy, or invoke an alternative route when the predefined algorithm is inadequate.

\textbf{Bounded responsibility and verification.} An agentic program is designed around a clearly defined scientific responsibility rather than open-ended research activity. Open-ended scientific tasks often admit multiple valid approaches and outcomes, making it difficult to define exhaustive criteria for correctness and verification protocols in advance~\cite{blade,human_reasoning3}. By bounding the responsibility, its scope, allowed tools, inputs, outputs, and scientific constraints can be specified in advance. This boundedness is important because it makes rigorous task-specific verification possible. Judgments made during execution should be checked whenever possible by deterministic calculations, physical constraints, consistency tests, or other task-specific gates. Not every judgment can be validated by such gates, but its computable consequences should be tested and the supporting evidence and judgment kept explicit and auditable. In this way, flexibility is allowed inside the program without sacrificing a well-defined standard for what constitutes an acceptable result.

\textbf{Maturation through episodes.} An agentic program need not begin as a fully autonomous system. Here, maturation refers to the progressive conversion of experience into persistent program artifacts such as doctrines, diagnostics, verification gates, and deterministic code~\cite{darwin,huang2025cascade}. In early development, human inspection and independent reviewer agents frequently expose previously unseen failure modes. These interventions do not remain isolated corrections, and each informative episode can be root-caused and codified into reusable instructions, verification gates, procedures, or deterministic algorithms. As those episodes accumulate, the program progressively incorporates knowledge that previously resided with human developers or emerged from reviewer feedback, reducing the need for repeated intervention. Furthermore, as recurring judgments are codified, maturation can also reduce the amount and complexity of LLM-based reasoning required for routine cases, shifting execution progressively toward deterministic components. Because codification itself can be incomplete or overly general, these artifacts should remain version-controlled, testable, and reversible as further experience accumulates.


\textbf{Complete delegation in production.} The ultimate goal of maturation is to preserve and extend the clear division of labor that has made conventional scientific software so powerful. Humans define the scientific objective, scope, and governing principles and improve the program when new failure modes are discovered. Routine execution, however, should belong to the program. This maturation resembles conventional software engineering, in which recurring issues and bugs are analyzed and fixed~\cite{conventional_software2, conventional_software}. In an agentic program, however, LLM-based judgment can participate in this process by interpreting new failure episodes and translating them into reusable program artifacts~\cite{darwin}, thereby reducing repeated human intervention within a clearly bounded scientific domain. Once sufficiently mature, an agentic program should operate without a human inside the production loop, either delivering a validated result or recognizing that a reliable result cannot be produced. This complete delegation distinguishes an agentic program from an interactive assistant or a human-supervised agentic workflow.

Modern agent harnesses make this form of software practical by providing both a development environment and a runtime for LLM-based judgment. By combining LLMs with tool execution, persistent context, reusable skills, subagents, and access to conventional scientific codes, platforms such as Claude Code~\cite{claude_code} and Codex~\cite{codex} allow judgment procedures to be developed, tested, versioned, and then executed as integral components of a scientific program. The harness is therefore not itself a defining property of an agentic program, but the infrastructure that enables its judgment-bearing components to be both developed and operated.

We remark that the concept of an agentic program differs in emphasis from much of the current work on agentic AI for science~\cite{huang2025cascade, zou2025agente, matsci_agent1, park2026robust, deng2026harnessing}. The individual elements described above are not necessarily new; rather, an agentic program is characterized by how these elements are brought together to enable the end-to-end delegation of a well-defined scientific task. Existing systems are often framed as assistants, workflow orchestrators, or increasingly general autonomous research agents, with the primary goal of expanding the range of scientific activities that an AI system can perform. An agentic program instead emphasizes complete delegation of a bounded scientific responsibility rather than maximal autonomy over the broadest possible range of tasks. Its scope is deliberately restricted so that the input, expected output, governing principles, and criteria for correctness can be defined sufficiently well to support systematic verification and repeated production.

\begin{figure}
\centering
  \includegraphics[width=9cm]{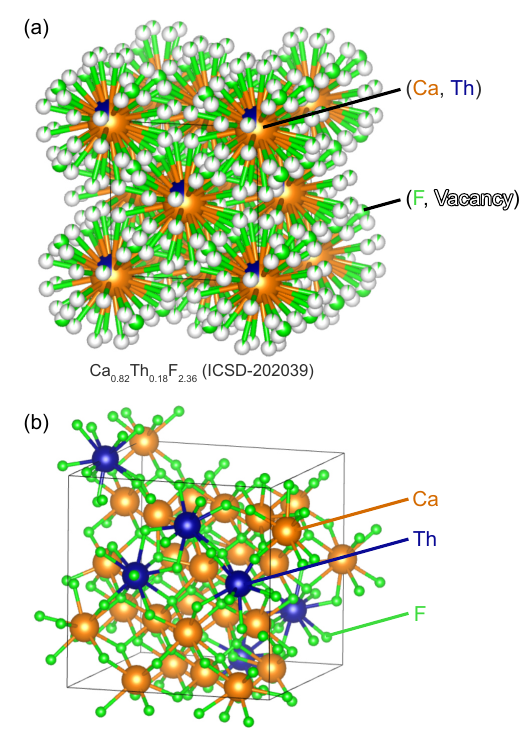}
  \caption{\textbf{Visualization of a disordered crystal showing complex behavior, Ca$_{1-x}$Th$_{x}$F$_{2+2x}$ with $x=0.18$.} (a) Crystal structure from the original CIF having partial occupancy. (b) A minimal atomistic representation (MAR) of (a) that follows chemical rules.}
  \label{fgr:disorder_occupation}
\end{figure}

\section{DeMARS: a prototype agentic program}

\textcolor{black}{To demonstrate how the four defining characteristics of agentic programs can be integrated in practice, we present DeMARS (De-averaged Minimal Atomistic Representation System) as a prototypical case study of an agentic program.} DeMARS is designed for a narrowly defined yet intrinsically judgment-intensive task: constructing atomistic models from experimentally reported disordered crystal structures given in a crystallographic information file (CIF). Additional details of DeMARS and its output artifacts are provided in Supplementary Notes~1--4. The program is also practically important, as approximately half of the entries in the Inorganic Crystal Structure Database (ICSD) are reported as disordered~\cite{disorder_in_icsd}. For compounds with only substitutional disorder, which account for $\sim$60\% of the disordered compounds~\cite{disorder_in_icsd}, a physically reasonable model can often be obtained by expanding the unit cell and distributing the partially occupied species over the corresponding sites. More complex disorder, however, cannot always be resolved by such straightforward enumeration. For example, in Ca$_{0.82}$Th$_{0.18}$F$_{2.36}$ (ICSD collection code 202039; Fig.~\ref{fgr:disorder_occupation}a), fluorine is distributed over several partially occupied normal and interstitial sites whose occupations are strongly coupled, making independent site-by-site enumeration impractical. The composition, however, reveals the underlying chemistry: replacing Ca$^{2+}$ by Th$^{4+}$ introduces two excess positive charges per Th and therefore requires two additional F$^-$ for compensation. This predicts an anion content of $2+2x=2.36$ at $x=0.18$, consistent with the refined composition~\cite{CaThF}. This reasoning suggests placing the additional F atoms at interstitial sites associated with Th (Fig.~\ref{fgr:disorder_occupation}b). The major difficulty is therefore not simply to enumerate configurations, but to determine what chemical model should be enumerated and which structural principles should constrain the search, a decision that requires scientific judgment.

\begin{figure*}
\centering
  \includegraphics[width=18cm]{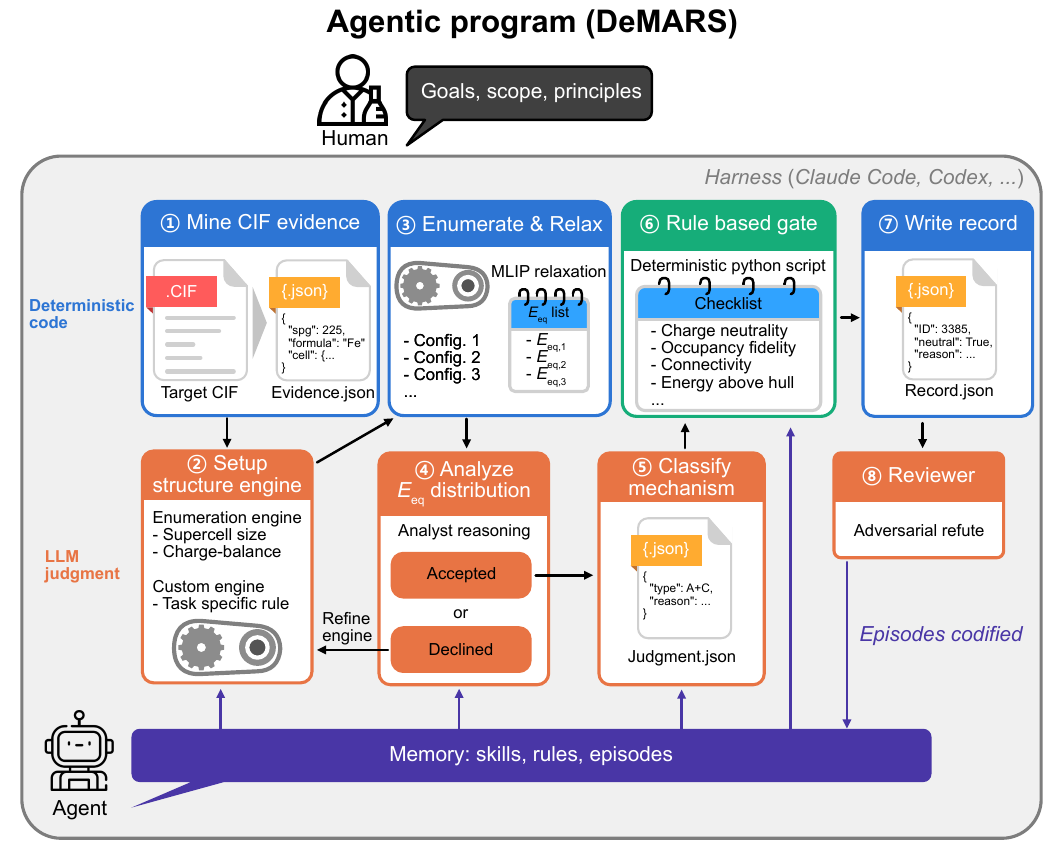}
  \caption{\textbf{Architecture of an agentic program illustrated through DeMARS.} Orange boxes represent the steps requiring LLM-based judgment, while blue and green boxes illustrate deterministic procedures by Python scripts. Processes colored in blue were developed from scratch for predefined purposes, while the green box indicates the verification gate containing numerous checklists that might require updates after learning episodes. Agents operating within the harness carry skills, rules and episodes as their memory (represented as purple box), which are continuously updated through human interaction by providing goals, scope and principles of the target (top) during maturation. $E_{\rm{eq}}$ indicates the equilibrium energy of enumerated crystals that can be obtained by structure optimization employing an MLIP.}
  \label{fgr:demars_arch}
\end{figure*}

Fig.~\ref{fgr:demars_arch} illustrates a general architectural pattern for agentic programs through the example of DeMARS, showing how deterministic code, LLM-based judgment, and persistent memory are integrated.
The architecture of DeMARS reflects the four characteristics described above: deterministic algorithms are combined with bounded LLM judgment, task-specific gates \textcolor{black}{test every output against predefined requirements}, recurring failures are incorporated through episodic maturation, and the resulting program is intended for end-to-end production without routine human intervention. Deterministic code (blue and green boxes) performs every operation whose procedure and acceptance criterion can be specified in advance: extracting the crystallographic evidence from the CIF into a structured bundle (\circled{1}), constructing supercells, enumerating candidate configurations, and relaxing them with a universal MLIP (\circled{3}), evaluating every deliverable against the verification gates, including charge neutrality, occupancy fidelity, and polyanion connectivity, whose outcomes are recorded for the reviewer, while also computing the energy above hull ($E_{\rm{hull}}$) of the final MAR as an interpretive metric rather than using it as a pass or fail criterion (\circled{6}), and assembling the final machine-readable record (\circled{7}). 

On the other hand, LLM-based judgment in DeMARS is introduced only at the underdetermined decision points. An LLM analyst operating within the harness reads the evidence bundle and converts its reading of the disorder into an executable specification for the enumeration engine, such as the supercell commensurate with the fractional occupancies, the sites to be merged or mutually excluded, the charge-balanced composition targets, or redirects the workflow when model repair (e.g., missing hydrogen) or a custom construction is required (\circled{2}); inspects the resulting energy distribution across the ensemble and decides whether to accept it or rerun the enumeration engine with a tightened specification (\circled{4}); and makes the mechanism classification (\circled{5}). A separate reviewer agent then adversarially reviews the finished record in a fresh context (\circled{8}). \textcolor{black}{Both agents were operated with Claude Sonnet 5 at high reasoning effort (see Supplementary Note 1).} The LLM therefore decides how the engine should be used when the appropriate procedure cannot be prescribed reliably beforehand.

\begin{figure*}
\centering
  \includegraphics[width=18cm]{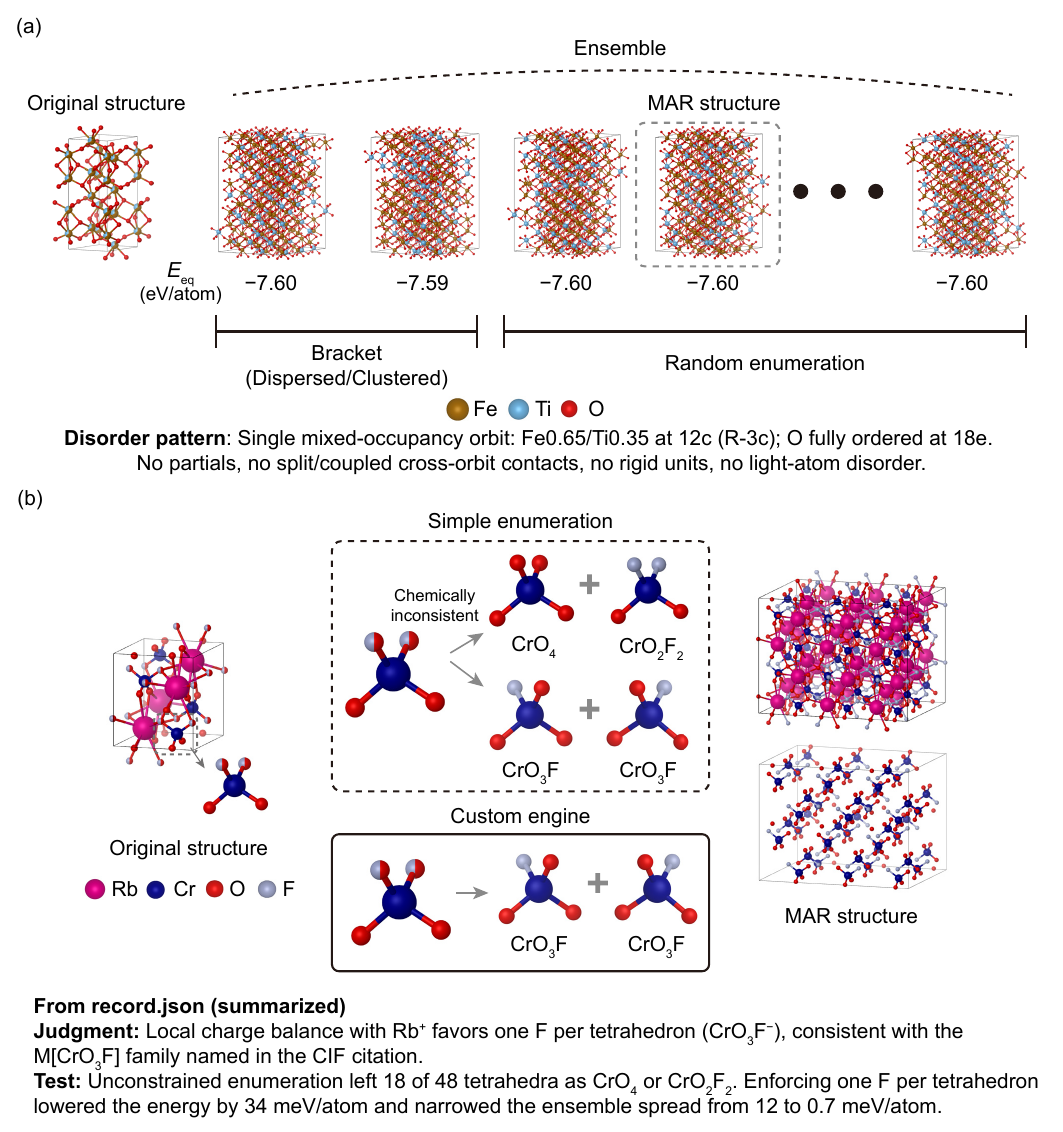}
  \caption{\textcolor{black}{\textbf{Execution of DeMARS on two disordered structures.} (a) ICSD 91597 (Fe$_{1.3}$Ti$_{0.7}$O$_{3}$), an entry with Fe/Ti substitutional disorder on a single site. The enumeration engine relaxes randomly decorated supercells together with dispersed and clustered configurations, which bracket the possible arrangements and are excluded from MAR selection by default. The narrow energy distribution of the random configurations (standard deviation 1.0~meV/atom) is consistent with the reported structure being the high-temperature disordered phase, measured in situ at 1373~K, and the lowest-energy random configuration is selected as the MAR. (b) ICSD 31704 (\ch{RbCrO3F}), in which each Cr tetrahedron contains two sites of a mixed O/F orbit. Simple enumeration of this orbit produces \ch{CrO4} and \ch{CrO2F2} units, whereas the custom engine preserves the \ch{CrO3F} unit in every tetrahedron. The MAR is shown in full (top) and with only the \ch{CrO3F} units (bottom). The text below summarizes the analyst's judgment recorded in record.json.}}
  \label{fgr:execution}
\end{figure*}

This architecture provides two structural safeguards for reliable judgment. First, the LLM operates between deterministic interfaces: code supplies the evidence it interprets, and code independently checks the consequences of its decisions. The model therefore neither generates the quantitative facts on which its judgment rests nor certifies its own output. These gates do not prove that a scientific interpretation is uniquely correct, but they test its computable consequences against predefined physical and structural requirements. Second, judgment is made falsifiable through computation. The analyst's specification is treated as a hypothesis about the disorder, whose physical consequences are tested by the engine; inconsistent structures, diagnostics, or energy distributions trigger refinement rather than acceptance. \textcolor{black}{Fig.~\ref{fgr:execution} illustrates how judgment can guide the search for candidate structures. For the Fe/Ti solid solution (ICSD 91597), generic random enumeration produces structures with similar energies, supporting its use without an additional constraint (Fig.~\ref{fgr:execution}a). For \ch{RbCrO3F}, however, randomly assigning O and F across the mixed orbit can generate tetrahedra with zero or two F atoms, even when the overall composition is correct. Using local charge balance and the \ch{CrO3F} family identified in the CIF citation, the analyst proposed that each tetrahedron contains one F and specified this hypothesis as a construction rule. The custom engine then generated structures that satisfy the rule, while randomly choosing which of the two mixed sites carries F (Fig.~\ref{fgr:execution}b and Supplementary Note 4). Compared with unconstrained enumeration, which produced \ch{CrO4} or \ch{CrO2F2} units in 18 of the 48 tetrahedra, the constraint lowered the energy of the final structure by 34~meV/atom. Thus, judgment helps determine which structures to generate, while computation provides evidence with which to assess that judgment.}

While the overall architecture of DeMARS remained stable, its components, including code, skills, doctrines, and verification procedures, were progressively refined through episodic experience within the Claude Code harness. Early versions required frequent human inspection as previously unseen forms of disorder exposed shortcomings in construction rules or verification. When a failure was identified, a brief human observation could trigger root-cause analysis by the orchestrating LLM, which then translated the insight into persistent program artifacts such as versioned instructions, diagnostics, procedures, or deterministic code. Reviewer agents were also used to expose overlooked failure modes, but did not modify the program themselves; updates were made between production runs by the orchestrating agent, with substantive doctrine changes subject to human approval. Across more than 800 disordered CIFs, recurring judgments were progressively codified into reusable rules and, where possible, deterministic algorithms. Thus, maturation progressively shifted recurring decisions from transient model-based judgment toward persistent and testable program artifacts. Several cases that initially required custom construction were later handled automatically by the improved engine. To enable explicit tracking of subsequent maturation cycles, we implemented a version management module within DeMARS (see Supplementary Note 2).


\textcolor{black}{One episode illustrates how a documented failure led to a persistent correction in the deterministic code (see Supplementary Note 2 and Fig.~S1). In \ch{[(CH3)2NH2]2K[Fe(CN)6]} (ICSD 239241), carbon occurs in two distinct chemical environments, as cyanide carbon bonded to one nitrogen atom and as methyl carbon bonded to four neighboring atoms. The original connectivity gate assigned a single coordination expectation to each element and incorrectly treated cyanide nitrogens, which serve as ligands, as independent coordination centers. These misclassifications resulted in 160 reported defects in a chemically correct structure. Following documentation of this failure as a maturation episode, the gate was revised to assign coordination expectations by chemical role. The revised gate reports no connectivity defects for this structure.}

Rapid maturation also depends on the ability to accumulate many relevant episodes at low computational cost. In atomistic materials problems, universal MLIPs provide this capability by allowing large numbers of candidate structures to be generated, relaxed, and compared far more efficiently than with first-principles calculations. DeMARS exploits this high-throughput experience through SevenNet~\cite{sevennet_omni, sevennet_nano}, using repeated evaluation of alternative disorder realizations to support judgment, verification, and subsequent maturation.

The consequence of this maturation is complete delegation within the defined scope of the program. Thus, the mature DeMARS can process randomly selected batches of disordered CIFs end-to-end without routine case-by-case human intervention, choosing between generic enumeration and custom construction, with model repair applied upstream of either where required. As a test, we randomly selected 100 disordered CIFs from ICSD (two to five elements, nontrivial site disorder, at most 40 listed sites, see Supplementary Note~5). DeMARS then processed them end-to-end without human intervention. Subsequent manual inspection found that 93 yielded chemically reasonable MARs consistent with the reported disorder, whereas seven did not meet the criteria for successful construction. Nevertheless, six of these seven still showed qualitative agreement with the corresponding original structures in the calculated X-ray diffraction (XRD) patterns (see Supplementary Notes 5 and 6 for detailed analyses). \textcolor{black}{To assess what analyst-directed construction adds beyond generic enumeration, we repeated the evaluation with the custom engine and analyst-side model repair disabled. The number of successful constructions decreased from 93 to 86, and every loss occurred among the 12 custom-engine cases (see Supplementary Note 8).} Complete delegation does not imply that every input must yield a successful structure; a reliable program should also be able to recognize when the available evidence and verification criteria are insufficient and decline to certify a result. What matters is that routine execution, including such decisions, no longer requires a human inside the production loop. At this stage, the agentic workflow begins to function as a scientific program in the conventional sense: a researcher delegates a defined task and receives either a validated result or an explicit failure.

\section{Challenges and Outlook}

Agentic programs also introduce challenges that are less prominent in conventional scientific software. In particular, LLM-based judgment is inherently sensitive to the context and employed model, so the same inputs may not always lead to the same decisions (see Supplementary Note 7). Hallucinations or missing physical or chemical considerations cannot always be caught by deterministic gates. In deployment, therefore, one needs careful versioning of models, skills, and episodic records, and continued auditing as the program or its underlying LLMs evolve. In this sense, the central challenge of agentic scientific software is not simply to make LLMs more capable, but to engineer the surrounding system so that their judgments remain inspectable, testable, and bounded.

\textcolor{black}{More capable frontier models may reduce the time and effort needed to mature task-specific agentic programs, but codifying experience into reusable rules and verification procedures should remain important. GPT-6 Astra, a frontier model from OpenAI at the time of this study, captured key chemical features even with a simple task-level prompt in Codex without any context but missed some details addressed by DeMARS (see Supplementary Note~9). This suggests that details such as distance constraints still need to be learned and verified through execution and experience, even when the broad chemistry is understood, to support consistent execution and complete delegation.}

The relative stability of much of the underlying physical and chemical knowledge may also make computational materials science particularly amenable to agentic programs, since many judgments depend less on rapidly changing external information than in some other domains.
Many problems in computational materials science have a structure that can be addressed by agentic programs: the input and desired deliverable are well defined, but reaching that deliverable involves intermediate decisions that cannot be fully prescribed in advance. Examples include reasoning-augmented crystal structure prediction, selecting and validating DFT protocols, constructing defect or interface models, and diagnosing and improving MLIPs. Rather than addressing these problems with a single general-purpose autonomous scientist, an alternative path may be to develop specialized agentic programs, each responsible for a bounded scientific task and matured through repeated experience within that domain. Such programs could then be composed with conventional simulation codes and with one another, much as scientific software packages are combined in present-day workflows. In this view, the future of computational materials science may lie in an ecosystem of reliable, specialized agentic programs to which bounded scientific responsibilities can be completely delegated. More broadly, the same paradigm may extend across science and engineering wherever well-defined technical responsibilities involve intermediate judgments that cannot be fully prescribed in advance.

\section{Data availability}
\textcolor{black}{The DeMARS outputs for the 100 hold-out test cases, the corresponding outputs of the custom-engine ablation, the execution example for ICSD 31704, and the ICSD collection codes of the entries used for maturation are available at \url{https://doi.org/10.5281/zenodo.22743157}. The input CIFs are not included because the ICSD is a licensed database. They can be retrieved from the ICSD using the collection codes.}

\section{Code availability}
The DeMARS source code, including the deterministic computational tools, agent definitions, and versioned skills used for model-based judgment, is available at \url{https://github.com/yslim1/DeMARS}.

\section{Acknowledgements}

This work was supported by Samsung Electronics Co., Ltd (IO260312-15844-01) and the Nano \& Material Technology Development Program through the National Research Foundation of Korea (NRF) funded by the Ministry of Science and ICT (RS-2026-25542918). The computations were carried out at the Korea Institute of Science and Technology Information (KISTI) National Supercomputing Center (KSC-2025-CRE-0284).

\section{Author contributions}

S.H. conceived the idea and developed the concept together with all authors.
Y.L., H.J., and S.H. developed the code base.
Y.L. and H.J. performed the experiments with assistance from J.K.$^1$ and J.K.$^2$.
Y.L., H.J., J.K.$^1$, J.K.$^2$, and S.H. prepared the manuscript.
All authors contributed to discussions and approved the paper.
(J.K.$^1$: Jaesun Kim, J.K.$^2$: Jisu Kim)

\section{Competing interests}
The authors declare no competing interests.

\bibliography{sn-bibliography}


\end{document}